\documentstyle[preprint,eqsecnum,aps]{revtex}

\begin{document}

\count255=\time\divide\count255 by 60 \xdef\hourmin{\number\count255}
  \multiply\count255 by-60\advance\count255 by\time
  \xdef\hourmin{\hourmin:\ifnum\count255<10 0\fi\the\count255}

\draft
\preprint{\vbox{\hbox{WM-98-108}\hbox{JLAB-THY-98-26}
}}

\title{Masses of Orbitally Excited Baryons in Large $N_c$ QCD}

\author{Carl E. Carlson$^\dagger$, Christopher D. Carone$^\dagger$,
Jos\'{e} L. Goity$^{\ddagger\diamond}$, and Richard
F. Lebed$^\diamond$}

\vskip 0.1in

\address{$^\dagger$Nuclear and Particle Theory Group, Department of
Physics, College of William and Mary, Williamsburg, VA 23187-8795\\
$^\ddagger$Department of Physics, Hampton University, Hampton, VA
23668 \\ $^\diamond$Jefferson Lab, 12000 Jefferson Avenue, Newport
News, VA 23606}

\vskip .1in
\date{July, 1998}
\vskip .1in

\maketitle

\begin{abstract}
We present the first phenomenological study of the masses of orbitally
excited baryons in large $N_c$ QCD\@.  Restricting here to the
nonstrange sector of the $\ell=1$ baryons, the $1/N_c$ expansion is
used to order and select a basis of effective operators that spans the
nine observables (seven masses and two mixing angles).  Fits are
performed using subsets of the complete set of nine operators,
including corrections up to $O(1/N_c)$ where leading order is $N_c^1$.
This study shows that the $1/N_c$ expansion provides an excellent
framework for analyzing the mass spectrum, and uncovers a new
hierarchy of operator contributions.
\end{abstract}

\thispagestyle{empty}

\newpage
\setcounter{page}{1}


\section{Introduction} \label{sec:intro}


        It appears that QCD admits a useful and elegant expansion in
powers of $1/N_c$, where $N_c$ is the number of colors\cite{tHooft}.
There are explicit rules that determine the order in $N_c$ of any
given Feynman diagram or matrix elements of any given operator.  One
can thus isolate first the leading contributions to the observable
under consideration and then systematically include contributions that
are proportional to higher powers of $1/N_c$.

        One may question whether $1/N_c = 1/3$ is small enough to
provide a valid phenomenological expansion parameter.  Experience
suggests that it is.  For the ground state baryons, the large $N_c$
approach has been used successfully to study SU(6) spin-flavor
symmetry\cite{DM,Jenk1,DJM1,CGO,Luty1},
masses\cite{DJM1,Jenk2,DJM2,JL}, magnetic
moments\cite{DJM1,DJM2,JM,Luty2,DDJM}, and axial current matrix
elements\cite{DM,DJM1,DJM2,DDJM}.

        The next natural step in this progression is studies of
excited baryons, in particular of the mixed symmetry, negative parity
{\bf 70}-plet of SU(6).  There have been studies of the
strong\cite{CGKM,PY2} and radiative\cite{CC} decays of these states
and of the structure of axial operator matrix elements\cite{PY1}.
Progress\cite{Goity} along the lines of the present work has already
been made with the first non-leading mass operators.

        In this paper, we study the relative order $1/N_c$ and
$1/N_c^2$ corrections to the masses of the nonstrange members of the
{\bf 70}-plet.  Qualitatively, we find that the large $N_c$ limit is
accurate for the {\bf 70}-plet, in the sense that all the mass
operators---if one wishes, all the mass terms in an effective
Hamiltonian---give contributions at or below the level estimated from
large $N_c$ considerations.  None are larger.  Quantitative detail is
added to this statement in Section III.

        Before proceeding, we make some comments on the nature of the
{\bf 70}-plet and the nomenclature we use.  We describe the state as a
symmetrized ``core'' of $(N_c-1)$ quarks in the ground state plus one
excited quark in a relative $P$ state.  The wave function is
antisymmetric in color and symmetric in SU(6) $\times$ O(3), where
SU(6) is the spin-flavor symmetry and O(3) is the rotation group.  We
use SU(6) to classify the large $N_c$ baryon states and the
transformation properties of the operators; however, we need not
assume that SU(6) is an exact symmetry.  In fact, while the leading
order contribution to the masses is of $O(N_c)$, SU(6) is broken at
$O(N_c^0)$ for our states.

        One can analyze the masses in the {\bf 70}-plet by expressing
the effective Hamiltonian as a sum of operators, one of which is
proportional to the identity and the rest are products of SU(6)
$\times$ O(3) generators times numerical coefficients.  Each operator
contributes to the mass at a definite order in $N_c$, which is
determined by rules delineated in Sec.~\ref{sec:melements}.

        The operator analysis for the excited baryons involves more
distinguishable generators than the ground state baryons, because of
the orbital angular momentum and because one quark is singled
out. This leads to a much larger collection of mass operators.  A list
is given in the next section, which shows that there is one operator
of $O(N_c)$, two operators of $O(1)$, and many of $O(1/N_c)$ or
smaller.

        Regarding the states, nonstrange mixed symmetry SU(6) states
with one quark singled out have total quark spin and isospin related
by $S=I$ or $I \pm 1$, with each of $S$ and $I$ in the range $1/2$ to
$N_c/2$.  (There is one exception: There are no doubly maximal mixed
symmetry $S=I=N_c/2$ states.)  Thinking of the states as a core of
$N_c-1$ ground state quarks with spin $S_c$ and isospin $I_c$ combined
with an excited quark with angular momentum $\ell =1$ leads to writing
the SU(6) $\times$ O(3) states, with the help of Clebsch-Gordan
coefficients, as

\begin{eqnarray}
\left| J J_3; I I_3 \, (\ell, \, S=I+\rho) \right\rangle &=&
  \sum_{m_\ell,m_1,\alpha_1,\eta}
                  \left(
                        \begin{array}{cc|c}
                             \ell & S   & J \\
                             m_\ell & m & J_3
                        \end{array}
                                          \right)
                  \left(
                        \begin{array}{cc|c}
                             S_c & 1/2   & S \\
                             m_1 & m_2   & m
                        \end{array}
                                          \right)
                  \left(
                        \begin{array}{cc|c}
                          I_c      & 1/2      & I \\
                          \alpha_1 & \alpha_2 & I_3
                        \end{array}
                                          \right)
      c_{\rho,\eta}
                         \nonumber \\[2ex]
&\times&
          \left|S_c=I_c=I+\eta/2;m_1,\alpha_1 \right>
\otimes
          \left|1/2;m_2,\alpha_2 \right>
\otimes
          \left|  \ell,m_\ell \right>
\end{eqnarray}
Here, the $m$'s are angular momentum projections, the $\alpha$'s are
isospin projections.  Note that $S_c=I_c \,$ since the (nonstrange)
core is symmetric in SU(6) indices, and we have written
$I_c=I+\eta/2$, where $\eta = \pm1$.  Note also that $\rho \equiv S-I
= \pm1,0$.  States with strangeness are defined analogously, except
that SU(3) Clebsch-Gordan coefficients appear in that case.  For
reasons of simplicity, we have restricted to nonstrange baryons in
this work.

        For $S=I\pm1$, the notation simplifies since $c_{\pm,\pm}=1$
and $c_{\pm,\mp}=0$.  For $S=I$,

\begin{eqnarray}
c_{0+} = + \sqrt{S \, (N_c+2(S+1)) \over N_c \, (2S+1) }
  \quad {\rm and} \quad
c_{0-} = - \sqrt{(S+1)(N_c-2S) \over N_c \, (2S+1) }.
\end{eqnarray}
(The orthogonal combination gives the totally symmetric SU(6) state.)
The explicit form of the state allows us to calculate analytically the
matrix elements for any operator for arbitrary $N_c$.

        For the physical case of $N_c=3$, the above expressions admit
the 7 nonstrange states of the {\bf 70}-plet.  Strictly speaking, the
label {\bf 70}-plet refers only to the mixed symmetry baryons
appearing at $N_c=3$; for large $N_c$, the representations tend to be
much larger. In the {\bf 70}-plet, the nonstrange states consist of
two isospin-$3/2$ states, $\Delta_{1/2}$ and $\Delta_{3/2}$, and five
isospin-$1/2$ states, $N_{1/2}$, $N^\prime_{1/2}$, $N_{3/2}$,
$N^\prime_{3/2}$, and $N^\prime_{5/2}$.  The subscript indicates total
baryon spin; unprimed states states have quark spin $1/2$ and primed
states have quark spin $3/2$.  In partial wave notation, the 2 deltas
and 5 nucleons are labeled $S_{31}, D_{33}, S_{11}$ (twice), $D_{13}$
(twice), and $D_{15}$, respectively.

        Section~\ref{sec:melements} explains the operator analysis and
gives matrix elements of a complete set of operators for the baryon
states.  Section~\ref{sec:results} contains our analysis of physical
masses and mixing angles.  Closing comments are made in
Sec.~\ref{sec:concl}.


\section{Operators and Matrix Elements} \label{sec:melements}


        Since the physically observed baryons are assigned to
irreducible representations of the symmetry group SU(6) $\times$ O(3),
it is natural to write all possible mass operators in terms of the
generators of this group.  O(3) is the group of spatial rotations and
is generated by the orbital angular momentum operator $\ell^i$, while
the spin-flavor group SU(6) has spin $S^i$, flavor $T^a$, and combined
spin-flavor $G^{ia}$ generators.  In the two-flavor case, these
operators are defined by
\begin{eqnarray}
S^i & \equiv & q^\dagger \left( \frac{\sigma^i}{2} \otimes \openone
\right) q , \nonumber \\ T^a & \equiv & q^\dagger \left( \openone
\otimes \frac{\tau^a}{2} \right) q ,\nonumber \\ G^{ia} & \equiv &
q^\dagger \left( \frac{\sigma^i}{2} \otimes \frac{\tau^a}{2} \right) q
,
\end{eqnarray}
where $\sigma^i$ and $\tau^a$ are the usual Pauli matrices.  The field
operators $q$, $q^\dagger$, which we term ``quarks,'' are not the
dynamical quarks, but rather eigenstates of the spin-flavor group such
that an appropriately symmetrized collection of $N_c$ of them have the
quantum numbers of the physical baryons. The collection of operators
constructed with these fields completely spans all possible physical
mass operators.  Only if the quarks are heavy can the fields $q$ be
identified with the dynamical valence quarks.

        The goal of the large $N_c$ analysis is to organize operators
by their effects on a given observable (in this case, the masses) in a
systematic expansion in powers of $N_c$.  Factors of $N_c$ originate
either as coefficients of operators in the Hamiltonian, or through
matrix elements of those operators.  For example, the unit operator
$\openone$ contributes at $O(N_c^1)$, since each quark contributes
coherently in the matrix element. The spin of the baryon $S^2$ is
known to contribute to the masses at $O(1/N_c)$\cite{Jenk2}, because
the matrix elements of $S^i$ are of $O(N_c^0)$ for baryons that have
spins of order unity as $N_c \to \infty$.  Similarly, matrix elements
of $T^a$ are $O(N_c^0)$ in the two-flavor case since the baryons
considered have isospin of $O(N_c^0)$, but the operator $G^{ia}$ has
matrix elements on this subset of states of $O(N_c^1)$.  This means
that the contributions of the $N_c$ quarks add incoherently in matrix
elements of the operator $S^i$ or $T^a$ but coherently for $G^{ia}$.
Note that matrix elements of a given operator are not necessarily
homogeneous in $N_c$; for example, values such as $N_c + 3$ can occur.

        In this work, the generators $S_c^i$, $T_c^a$, $G_c^{ia}$ are
reserved to mean those acting upon the core, while separate SU(6)
generators $s^i$, $t^a$, and $g^{ia}$ are defined for the single
excited-quark system. Including the operator $\ell^i$ completes the
list of building blocks for the necessary mass operators.  Strictly
speaking, the naive symmetry group for this operator basis is SU(6)
$\times$ SU(6) $\times$ O(3), although actually only the diagonal
subgroup SU(6) $\times$ O(3) truly acts on the baryon states.  Given
this enlarged notation, it is possible to construct a large number of
operators; however, many are linearly dependent and may be discarded.

        Since the excited system consists of only one quark, at most
one generator from among $s$, $t$, or $g$ appears in any operator.  A
similar but stronger statement may be made for the core: Since we
ultimately perform phenomenological analysis on cores with two quarks,
at most two of the set $\{ S_c, T_c, G_c \}$ are needed.  However,
operator reduction rules exist\cite{DJM2} that significantly reduce
the number of core operators that must be considered.  Finally, since
we are interested here in $\ell=1$ baryons, only operator combinations
of $\Delta \ell = 0,1,2$ need be considered. Hence, only up to two
factors of $\ell^i$ are required. Indeed, when two $\ell$'s appear, it
is convenient to use $\ell^{(2)}_{ij}$, the rank two tensor
combination of $\ell^i$ and $\ell^j$ given by
\begin{equation}
\ell^{(2)}_{ij} = \frac{1}{2} \{ \ell_i , \ell_j \} - \frac{\ell^2}{3}
\delta_{ij} .
\end{equation}

        The explicit power of $N_c$ for a given operator is determined
by using the usual large $N_c$ counting of $1/\sqrt{N_c}$ for each
quark-quark-gluon coupling, and considering the minimal number of
exchanged gluons necessary to generate a given operator.  To be
specific, we decompose an $n$-body operator ${\cal O}$ as follows:
\begin{equation}
{\cal O} = X_* \prod_{i=1}^{n-1} X_i ,
\end{equation}
where $X_*$ represents all operators acting on the excited quark,
including factors of $\ell^i$, and each $X_i$ represents an SU(6)
generator acting on the core.  The physical realization of such an
operator requires exchanging a minimum of $n-1$ gluons between
different quarks, leading to a suppression of $1/N_c^{n-1}$, which we
henceforth include in the definition of the operators.  If $X_* =
\openone$, the result is maintained as written if $n$ rather than
$n-1$ operators act on the core.

        One then considers each core operator $X_i$ to determine
whether its matrix elements are coherent for the baryon states under
consideration; a factor of $N_c$ is included for each coherent
operator.  Thus, the full large $N_c$ counting of the matrix element
is $O(N_c^{1-n+m})$, where $m$ is the number of coherent $X_i$.  In
the two-flavor case, only $G_c^{ia}$ has coherent matrix elements.
The order of the matrix element thus obtained determines whether or
not we retain the operator in computing a given process to a desired
order in the $1/N_c$ expansion.

        With this counting, one finds 22 potentially independent,
time-reversal even, isosinglet operators: One ($\openone$) with matrix
element of $O(N_c^1)$, two at $O(N_c^0)$, ten at $O(1/N_c)$, and nine
at $O(1/N_c^2)$ or higher.  This counting does not fully take into
account numerous relations between the matrix elements of the
operators evaluated on the nonstrange $\ell=1$ baryons.  One reduction
that has been included uses the observation that, for the nonstrange
mixed symmetry states,
\begin{equation}
\frac{1}{N_c} \left< gG_c \right> = -\frac{N_c+1}{16N_c} + \delta_{S,I}
\frac{I(I+1)}{2N_c^2} .
\end{equation}
This operator naively produces $O(N_c^0)$ matrix elements, but both
the $O(N_c^0)$ and $O(1/N_c)$ parts are the same for all baryons in
the multiplet, and thus may be absorbed into matrix elements of
$\openone$, with the remainder being demoted to $O(1/N_c^2)$.
Similarly, both $\left< \ell s \right>$ and $\left< \ell tG_c \right>$
are $O(N_c^0)$, but it may be observed that $\left< \ell s +4\ell
tG_c/(N_c+1) \right>$ is $O(1/N_c)$, so only $\left< \ell s \right>$
truly represents an independent $O(N_c^0)$ operator.  The full set of
operator reductions will be presented in a future publication
\cite{CCGL}.

        In any case, there are only 9 observables (masses and mixing
angles) in the nonstrange $\ell=1$ system, and so only 9 independent
operators are required.  An independent basis is presented in
Table~\ref{matel}: All 3 occurring up to $O(N_c^0)$ (first considered
in Ref.~\cite{Goity}), and a selection of 6 at $O(1/N_c)$ whose matrix
elements, when combined with the first 3, are seen to be independent
for $N_c=3$.  With these operators denoted by ${\cal O}_1, {\cal O}_2,
\ldots , {\cal O}_9$, respectively, the nine independent mass matrix
elements are given by
\begin{equation}
M_j = \sum_{i=1}^9 c_i \langle {\cal O}_i \rangle_j \  \,\,\,\,
      (j=1 \ldots 9) .
\end{equation}
The coefficients $c_i$ are independent of $N_c$ at leading order,
given our choice of operator normalization.  These operator
coefficients encapsulate all unknown strong interaction physics
unspecified by the large $N_c$ spin-flavor analysis.  In
Table~\ref{matel} we present the matrix elements $\langle {\cal O}_i
\rangle_j$.  Explicit spin and flavor indices are suppressed when
their contraction is unambiguous.


\section{Results} \label{sec:results}


        In addition to the nonstrange mixed symmetry states defined in
Sec.~\ref{sec:intro}, two mixing angles are necessary to specify the
$S=1/2$ and $S=3/2$ nucleon mass eigenstates.  We define
\begin{equation}
\left[\begin{array}{c} N(1535) \\ N(1650) \end{array} \right] =
\left[\begin{array}{cc}  \cos\theta_{N1} & \sin\theta_{N1} \\
                       -\sin\theta_{N1} & \cos\theta_{N1}
\end{array}\right]
\left[\begin{array}{c} N_{1/2} \\ N^\prime_{1/2}\end{array} \right]
\end{equation}
and
\begin{equation}
\left[\begin{array}{c} N(1520) \\ N(1700) \end{array} \right] =
\left[\begin{array}{cc}  \cos\theta_{N3} & \sin\theta_{N3} \\
                       -\sin\theta_{N3} & \cos\theta_{N3}
\end{array}\right]
\left[\begin{array}{c} N_{3/2} \\ N^\prime_{3/2}\end{array} \right]
\,\,\, ,
\label{eq:tpt}
\end{equation}
as in Ref.~\cite{CGKM}.  The mass eigenvalues and mixing angles can be
expressed in terms of the coefficients $c_i$ of the operators
presented in the previous section.

        Since we have found an operator basis that completely spans
the $9$-dimensional space of observables, we can solve for the $c_i$
given the experimental data.  For each baryon mass, we assume that the
central value corresponds to the midpoint of the mass range quoted in
the {\it Review of Particle Properties}~\cite{RPP}; we take the one
standard deviation error as half of the stated range.  To determine
the off-diagonal mass matrix elements, we use the mixing angles
extracted from the analysis of strong decays given in
Ref.~\cite{CGKM}, $\theta_{N1}=0.61\pm 0.09$ and $\theta_{N3}= 3.04
\pm 0.15$.  These values are consistent with those obtained in
\cite{CC} from radiative decays.  Solving for the operator
coefficients, we obtain the values shown in Table~\ref{csolve}.

        Naively, one expects the $c_i$ to be of comparable size. Using
the value of $c_1$ as a point of comparison, it is clear that there
are no operators with anomalously large coefficients.  Thus, we find
no conflict with the naive $1/N_c$ power counting rules.  It is
interesting that a number of the operators appear to be unimportant in
describing the experimental data (presumably due to the underlying
dynamics).  For example, of the two operators that contribute to the
masses at ${\cal O}(1)$, the operator ${\cal O}_2 = \ell s$ has a
coefficient which is suppressed relative to ${\cal O}_3 =
\ell^{(2)}gG_c /N_c$ by more than factor of $10$; in effect, this
operator is no more important than a typical $O(1/N_c^2)$ correction.
Of the operators that contribute to the masses at order $1/N_c$, only
the operator ${\cal O}_6 = S_c^2/N_c$ contributes as much as one would
expect, with the next largest corrections coming from the operators
${\cal O}_4 = \ell s+4\ell tG_c /(N_c+1)$ and ${\cal O}_5 = \ell
S_c/N_c$.

        Using these observations, we can attempt to fit the data using
judiciously chosen subsets of the original 9 operators.  We fit to the
seven mass eigenvalues as well as the two mixing angles $\theta_{N1}$
and $\theta_{N3}$. The operator set we consider first are ${\cal
O}_1$, ${\cal O}_2$, and ${\cal O}_3$; these yield mass predictions
accurate to order $1$ in the $1/N_c$ expansion, and thus present the
first nontrivial spin-flavor symmetry-breaking corrections. We show
the result of this fit in Table~\ref{lowest}.  In short, the lowest
order operators fail in reproducing the experimental data.  Notably,
the mixing angles are far off the mark, and the $J=1/2$ $\Delta$ state
is predicted to be heavier than the $J=3/2$ $\Delta$ state. The
difficulty in obtaining a good fit from a leading order analysis is an
outcome that perhaps could have been anticipated: Naively, one might
expect that the lowest nontrivial mass corrections are roughly a
factor of $N_c=3$ smaller than the mean baryon mass, or approximately
$500$ MeV.  The largest splitting within our set of seven baryons is
$\approx180$ MeV, for example, in the case of the N(1700)-N(1520) mass
difference. Thus, we might have concluded {\it {a} priori} that
$1/N_c$ corrections are necessary in order to reproduce the detailed
features of the mass spectrum.

        In the remaining fits, we include $1/N_c$ corrections.  The
$6$ parameter fit shown in Table~\ref{6param} includes all the
subleading operators that appear to be significant in
Table~\ref{csolve}; the operators included are ${\cal O}_1$ through
${\cal O}_6$.  The resulting fit is in extremely good agreement with
the experimental data, with no predicted mass more than 0.4 standard
deviations from the corresponding experimental central value, and a
$\chi^2$ per degree of freedom of $0.1$.

        More strikingly, Table~\ref{csolve} implies that we can
discard additional operators and still obtain a reasonable fit.
Notice that the smallness of the coefficients $c_2$, $c_4$, and $c_5$
renders the corresponding operators numerically unimportant, and thus
they can be neglected if we are only interested in working to order
$1/N_c$.  In Table~\ref{3param} we give a fit retaining the remaining
three operators, ${\cal O}_1$, ${\cal O}_3$, and ${\cal O}_6$.  The
$\chi^2$ per degree of freedom for this fit is $1.87$, which is not
bad considering that have only included two nontrivial operators.
Notice that the particular operator choice for this fit leads to a
degeneracy between the $\Delta_{1/2}$ and $\Delta_{3/2}$ which is
lifted by the corrections that we have discarded.  We do not display
the fits corresponding to all possible choices for the subleading
operators.  It suffices to point out that these additional fits
interpolate between the 6 and 3 parameter fits that we have presented
in Tables~\ref{6param} and \ref{3param}. One may also fit to the mass
eigenvalues and predict the mixing angles.  These fits are not
qualitatively different from the ones given here, and will be
presented in a longer publication \cite{CCGL}.  The crucial
observation is that the subleading operator ${\cal O}_6 = S_c^2/N_c$
is the most significant ingredient in taking us from the poor fit
shown in Table~\ref{lowest} to the good fits in Tables~\ref{6param}
and \ref{3param}.

\section{Conclusions} \label{sec:concl}

        The value of the large $N_c$ approach to baryon phenomenology
is that it provides an organizing principle in constructing the baryon
effective field theory.  Studies of the excited baryon mass spectrum
in the formative days of SU(6) found numerous operators\cite{heroes},
but in that period there was no organizing principle available to
select among them.  Beginning with a complete operator basis that
spans the space of any desired set of observables, large $N_c$ power
counting rules tell us which operators may be discarded if we wish to
obtain predictions to a desired level of accuracy.  In this sense, our
results for the masses and mixing angles of the nonstrange $\ell =1$
baryons are completely consistent with the large $N_c$ picture. We
find no operator with a coefficient that is larger than what one would
expect from the naive large $N_c$ power counting rules.

        More interesting, however, is that at any given order in
$1/N_c$, only some of the operators are of phenomenological relevance.
The $O(N_c^0)$ coefficients $c_i$ that we have defined in our
effective theory parametrize the long-distance physics that we cannot
calculate.  If we had found that these coefficients were of comparable
size, we might have concluded that large $N_c$ counting arguments
alone are sufficient to explain the detailed features of the mass
spectrum.  Quite the contrary, we find that only a few of our original
set of operators are needed to reproduce the experimental data, most
notably, the operators $S_c^2$ and $\ell^{(2)}gG_c$.  It is tempting
to speculate that the importance of the operator $S_c^2$ can be
understood by considering the explicit nonrelativistic reduction of a
one-gluon exchange interaction; the second operator, however, has a
nontrivial flavor structure that does not correspond to the usual
tensor interaction $\ell^{(2)}s S_c$ that one derives in this
approach~\cite{glozman}. A fit analogous to that of Table~\ref{3param}
replacing $\ell^{(2)}gG_c/N_c$ with $\ell^{(2)}s S_c/N_c$ increases
the $\chi^2$ per degree of freedom from 1.87 to 2.46.  Why the
underlying dynamics should prefer these operators is a much more
difficult question which goes beyond what can be addressed in the
effective field theory approach.

\begin{center}
{\bf Acknowledgments}
\end{center}
CEC thanks the National Science Foundation for support under Grant
PHY-9600415; CDC thanks the NSF for support under Grant PHY-9800741;
and JLG similarly thanks the NSF for support under Grant HRD-9633750.
Both JLG and RFL thank the Department of Energy for support under
Contract DE-AC05-84ER40150.


%
%

\begin{table}

\begin{tabular}{l||cc|cc|cc|cc|c}

& $\langle \openone \rangle$ && $\langle \ell s \rangle$ &&
$\frac{1}{N_c} \langle \ell^{(2)} gG_c \rangle$ && $\langle \ell s +
\frac{4}{N_c+1} \ell tG_c \rangle$ && $ \frac{1}{N_c} \langle \ell S_c
\rangle$ \\ \hline\hline

$N_{1/2}$ & $N_c$ && $-\frac{1}{3N_c} (2N_c-3)$ && 0 &&
$+\frac{2}{N_c+1}$ && $-\frac{1}{3N_c^2} (N_c+3)$ \\ \hline

$N^\prime_{1/2}$ & $N_c$ && $-\frac 5 6$ && $-\frac{5}{48N_c} (N_c+1)$
&& 0 && $-\frac{5}{3N_c}$ \\ \hline

$N^\prime_{1/2} \,$-$N_{1/2}$ & 0 &&
$-\frac{1}{3}\sqrt{\frac{N_c+3}{2N_c}}$ && $-\frac{5}{48N_c}
\sqrt{\frac{N_c+3}{2N_c}} (2N_c-1)$ && $-\frac{1}{N_c+1}
\sqrt{\frac{N_c+3}{2N_c}}$ &&
$+\frac{1}{3N_c}\sqrt{\frac{N_c+3}{2N_c}}$ \\ \hline

$N_{3/2}$ & $N_c$ && $+\frac{1}{6N_c} (2N_c-3)$ && 0 &&
$-\frac{1}{N_c+1}$ && $+\frac{1}{6N_c^2} (N_c+3)$ \\ \hline

$N^\prime_{3/2}$ & $N_c$ && $-\frac{1}{3}$ && $+\frac{1}{12N_c}
(N_c+1)$ && 0 && $-\frac{2}{3N_c}$ \\ \hline

$N^\prime_{3/2} \,$-$N_{3/2}$ & 0 &&
$-\frac{1}{6}\sqrt{\frac{5(N_c+3)}{N_c}}$ && $+\frac{1}{96N_c}
\sqrt{\frac{5(N_c+3)}{N_c}} (2N_c-1)$ && $-\frac{1}{2(N_c+1)}
\sqrt{\frac{5(N_c+3)}{N_c}}$ &&
$+\frac{1}{6N_c}\sqrt{\frac{5(N_c+3)}{N_c}}$ \\ \hline

$N^\prime_{5/2}$ & $N_c$ && $+\frac 1 2$ && $-\frac{1}{48N_c} (N_c+1)$
&& 0 && $+\frac{1}{N_c}$ \\ \hline

$\Delta_{1/2}$ & $N_c$ && $+\frac 1 3$ && 0 && 0 && $-\frac{4}{3N_c}$
\\ \hline

$\Delta_{3/2}$ & $N_c$ && $-\frac 1 6$ && 0 && 0 && $+\frac{2}{3N_c}$

\end{tabular}

\vskip 2ex

\begin{tabular}{l||cc|cc|cc|c}

& $\frac{1}{N_c} \langle S_c^2 \rangle$ && $\frac{1}{N_c} \langle tT_c
\rangle$ && $\frac{1}{N_c} \langle \ell^{(2)} sS_c \rangle$ &&
$\frac{1}{N_c^2} \langle \ell^i g^{ia} \{ S_c^j,G_c^{ia} \} \rangle$
\\ \hline\hline

$N_{1/2}$ & $+\frac{1}{2N_c^2} (N_c+3)$ && $-\frac{1}{4N_c^2} (N_c+3)$
&& 0 && $-\frac{1}{24N_c^3} (5N_c-1)(N_c+3)$ \\ \hline

$N^\prime_{1/2}$ & $+\frac{2}{N_c}$ && $-\frac{1}{N_c}$ &&
$+\frac{5}{6N_c}$ && $+\frac{5}{24N_c^2} (N_c+1)$ \\ \hline

$N^\prime_{1/2} \,$-$N_{1/2}$ & 0 && 0 && $+\frac{5}{12N_c}
\sqrt{\frac{N_c+3}{2N_c}}$ && $+\frac{1}{24N_c^2}
\sqrt{\frac{N_c+3}{2N_c}} (2N_c-1)$ \\ \hline

$N_{3/2}$ & $+\frac{1}{2N_c^2} (N_c+3)$ && $-\frac{1}{4N_c^2} (N_c+3)$
&& 0 && $+\frac{1}{48N_c^3} (5N_c-1)(N_c+3)$ \\ \hline

$N^\prime_{3/2}$ & $+\frac{2}{N_c}$ && $-\frac{1}{N_c}$ &&
$-\frac{2}{3N_c}$ && $+\frac{1}{12N_c^2} (N_c+1)$ \\ \hline

$N^\prime_{3/2} \,$-$N_{3/2}$ & 0 && 0 && $-\frac{1}{24N_c}
\sqrt{\frac{5(N_c+3)}{N_c}}$ && $+\frac{1}{48N_c^2}
\sqrt{\frac{5(N_c+3)}{N_c}} (2N_c-1)$ \\ \hline

$N^\prime_{5/2}$ & $+\frac{2}{N_c}$ && $-\frac{1}{N_c}$ &&
$+\frac{1}{6N_c}$ && $-\frac{1}{8N_c^2} (N_c+1)$ \\ \hline

$\Delta_{1/2}$ & $+\frac{2}{N_c}$ && $+\frac{1}{2N_c}$ && 0 &&
$+\frac{1}{6N_c^2}(N_c+1)$ \\ \hline

$\Delta_{3/2}$ & $+\frac{2}{N_c}$ && $+\frac{1}{2N_c}$ && 0 &&
$-\frac{1}{12N_c^2}(N_c+1)$

\end{tabular}

\caption{Matrix elements $\langle {\cal O}_i \rangle_j$ of 9
operators, labeled as ${\cal O}_1, {\cal O}_2, \ldots, {\cal O}_9$,
respectively, that are linearly independent for $N_c=3$.  The third
and sixth rows correspond to off-diagonal matrix
elements.\label{matel}}

\end{table}


\begin{table}
\begin{tabular}{cc|ccc|cccccc}
$c_1$ && $c_2$ & $c_3$ && $c_4$ & $c_5$ & $c_6$ & $c_7$ & $c_8$ &
$c_9$ \\ \hline
$+0.470$ && $-0.036$ & $+0.369$ && $+0.089$ & $+0.087$ & $+0.418$ &
$+0.040$ & $+0.048$ & $+0.012$ \\
$\pm 0.017$ && $\pm 0.041$ & $\pm 0.208$ && $\pm 0.203$ & $\pm 0.157$
& $\pm 0.085$ & $\pm 0.074$ & $\pm 0.172$ & $\pm 0.673$ \\
\end{tabular}
\caption{Operator coefficients in GeV, assuming the complete set of
Table~\protect\ref{matel}. The vertical divisions separate operators
whose contributions to the baryon masses are of orders $N_c^1$,
$N_c^0$ and $N_c^{-1}$, respectively.\label{csolve}}
\end{table}


\begin{table}
\begin{tabular}{cccc|ccc}
\multicolumn{7}{c}{Parameters (GeV): $c_1=0.542\pm 0.002$,
$c_2=0.093\pm 0.008$, $c_3=-0.335\pm 0.039 $} \\
\hline\hline
               & Fit (MeV) & Exp.\ (MeV) && & Fit (MeV) & Exp.\ (MeV)
\\ \hline
$\Delta(1700)$ & $1610$ & $1720\pm 50$ &&$N(1520)$ & $1520$ & $1523\pm
8$ \\
$\Delta(1620)$ & $1657$ & $1645\pm 30$ &&$N(1535)$ & $1568$ & $1538\pm
18$\\
$ N(1675) $ & $1682$ & $1678\pm 8$ && $\theta_{N1}$ & $0.79$ &
$0.61\pm 0.09$ \\
$ N(1700) $ & $1679$ & $1700\pm 50$ && $\theta_{N3}$ &$2.63$ &
$3.04\pm 0.15$ \\
$N(1650)$   &     $1622$ & $1660\pm 20$ &&  & &
\end{tabular}
\caption{Three parameter fit using operators ${\cal O}_{1,2,3}$,
giving $\chi^2/{\rm d.o.f.}=23.33/6=3.89$. The operators included
formally yield the lowest order nontrivial contributions to the masses
in the $1/N_c$ expansion.}
\label{lowest}
\end{table}


\begin{table}
\begin{tabular}{cccc|ccc}
\multicolumn{7}{c}{Parameters (GeV): $c_1=0.468\pm 0.005$,
$c_2=-0.032\pm 0.045$, $c_3=0.327\pm 0.093$} \\
\multicolumn{7}{c}{$c_4=0.081\pm 0.027$, $c_5=0.071\pm 0.042$,
$c_6=0.413\pm 0.044$}\\  \hline\hline
               & Fit (MeV) & Exp.\ (MeV) && & Fit (MeV) & Exp.\ (MeV)
\\ \hline
$\Delta(1700)$ & $1701$ & $1720\pm 50$ &&$N(1520)$ & $1523$ & $1523\pm
8$ \\
$\Delta(1620)$ & $1637$ & $1645\pm 30$ &&$N(1535)$ & $1537$ & $1538\pm
18$\\
$ N(1675) $ & $1678$ & $1678\pm 8$ && $\theta_{N1}$ & $0.60$ &
$0.61\pm 0.09$ \\
$ N(1700) $ & $1712$ & $1700\pm 50$ && $\theta_{N3}$ &$3.06$& $3.04\pm
0.15$ \\
$N(1650)$   &     $1662$ & $1660\pm 20$ &&  & &
\end{tabular}
\caption{Six parameter fit using operators ${\cal O}_{1,\cdots,6}$,
giving $\chi^2/{\rm d.o.f.}=0.31/3=0.10$.}
\label{6param}
\end{table}


\begin{table}
\begin{tabular}{cccc|ccc}
\multicolumn{7}{c}{Parameters (GeV): $c_1=0.461\pm 0.005$,
$c_3=0.360\pm 0.059$, $c_6=0.453\pm 0.030$} \\ \hline\hline
               & Fit (MeV) & Exp.\ (MeV) && & Fit (MeV) & Exp.\ (MeV)
\\ \hline
$\Delta(1700)$ & $1683$ & $1720\pm 50$ &&$N(1520)$ & $1530$ & $1523\pm
8$ \\
$\Delta(1620)$ & $1683$ & $1645\pm 30$ &&$N(1535)$ & $1503$ & $1538\pm
18$\\
$ N(1675) $ & $1673$ & $1678\pm 8$ && $\theta_{N1}$ & $0.45$ &
$0.61\pm 0.09$ \\
$ N(1700) $ & $1725$ & $1700\pm 50$ && $\theta_{N3}$ &$3.04$& $3.04\pm
0.15$ \\
$N(1650)$   &     $1663$ & $1660\pm 20$ &&  & &
\end{tabular}
\caption{Three parameter fit using operators ${\cal O}_1$, ${\cal
O}_3$, and ${\cal O}_6$, giving $\chi^2/{\rm d.o.f.}=$
$11.19/6=1.87$.}
\label{3param}
\end{table}

\end{document}